\documentclass[12pt,fleqn]{article}
\usepackage{psfig,float}
\begin{document}
\begin{center}
{\Large{\bf A model for $\pi\Delta$ electroproduction}}
\end{center}
\begin{center}
{\Large{\bf on the proton}}
\end{center}

\vspace{1cm}

\begin{center}
{\large{ J.C. Nacher and E. Oset}}
\end{center}

\vspace{0.4cm}
\begin{center}
{\it Departamento de F\'{\i}sica Te\'orica and IFIC}
\end{center}
\begin{center} 
{\it Centro Mixto Universidad de Valencia-CSIC}
\end{center}
\begin{center}
{\it 46100 Burjassot (Valencia)}
\end{center}
\begin{center}
{\it Spain.}       
\end{center}

\vspace{3cm}

\begin{abstract}
{\small{We have extended a model for the $\gamma N\rightarrow \pi\pi N$
reaction to virtual photons and selected the diagrams which have a $\Delta$
in the final state. With this model we have evaluated cross sections for the
virtual photon cross section as a function of $Q^2$ for different energies.
The
agreement found with the $\gamma_v p\rightarrow\Delta^0\pi^+$ and  
$\gamma_v p\rightarrow\Delta^{++}\pi^-$ reactions is good. The sensitivity of 
the results to $N\Delta$ transition form factors is also studied. The present
reaction, selecting a particular final state, is an extra test for models of 
the 
$\gamma_{v} N\rightarrow \pi\pi N$ amplitude. The experimental measurement of the different isospin channels
 for this reaction are encouraged as a means to unravel the dynamics of the
 two pions photoproduction processes.}}
\end{abstract}
\newpage
\newpage
\section{Introduction}

The $\gamma N\rightarrow \pi\pi N$ reaction in nuclei has captured some
attention
recently and has proved to be a source of information on several aspects of 
resonance formation and decay as well as a test for chiral perturbation theory
at low energies. A model for the $\gamma p\rightarrow\pi^+\pi^- p$ reaction was
developed in [1] containing 67 Feynman diagrams by means of which a good 
reproduction of the cross section was found up to about $E_\gamma\simeq 1$ 
GeV.

A more reduced set of diagrams, with 20 terms , was found sufficient to describe
the reaction up to $E_\gamma\simeq 800$ MeV [2] where the Mainz 
experiments are done [3,4,5].

In the same work [2] the model was extended to the other charge channels of 
($\gamma,2\pi$) and the agreement with the experiment is good except for the 
channel $\gamma p\rightarrow\pi^+\pi^0 n$ where the theoretical cross section 
underestimates the experiment in about 40 per cent.

Other models have also been recently proposed. The model of ref. [6] contains the
dominant terms of [1] and includes some extra resonances in order to make predictions
at high energies. A prescription to approximately unitarize the model, of relevance at
high energies, is also proposed. Revision of this work is under 
consideration [7],
so more about it should be known in the future. 

The model of [8] has fewer diagrams than the one of [1,2] but introduces the 
$N^\ast(1520)\rightarrow N \rho$ decay mode. By fitting a few parameters to 
$(\gamma, \pi \pi)$ data the cross sections are reproduced,
including the $\gamma p\rightarrow\pi^+\pi^0 n$ reaction where the model of [2] fails.
The $(\gamma,\pi^0 \pi^0)$ channel is somewhat underpredicted.
The model of [8] fails to reproduce some invariant mass distributions where the model of
[1] shows no problems. A different version of the model of [8] is given 
in [9], where
the parameters of the model are changed in order to reproduce the mass distribution, 
without  spoiling the cross sections. One of the problems in the fit of [9] is that the
range parameter of the $\rho$ coupling to baryons is very small, around 200 MeV, which 
would not be easily accomodated in other areas of the $\rho$ phenomenology, like the
isovector $\pi N$ $s-wave$ scattering amplitude.

On the other hand the model of [2] has no free parameters. All input is obtained from known
properties of resonances and their decay, with some unknown sign borrowed from quark models.
 With all its global success in the different isospin channels, the persistance of the 
discrepancy in the $\gamma p\rightarrow\pi^+\pi^0 n$ channel indicates that 
further work is needed. The constraints provided by the data
in two pion electroproduction should be useful to further test present models and learn
more about the dynamics of the two pion production.

The ($\gamma$,$2\pi$) reaction has also been used to test chiral perturbation 
theory. The threshold region is investigated in [10,11] with some discrepancies 
in the results which are commented in [12,13]. The one loop corrections are 
shown to be relevant in the $\gamma p\rightarrow\pi^0\pi^0 p$ reaction close 
to threshold [11]. In addition the $N^\ast$(1440) excitation is shown to be 
very important at threshold in [2].

Another interesting information obtained from the $\gamma p\rightarrow
\pi^+\pi^- p$ reaction is information about the $N^\ast$(1520) decay into 
$\Delta\pi$. Indeed, the photoexcitation of the $N^\ast$(1520) followed by the 
decay into $\Delta\pi$ is a mechanism which interferes with the dominant 
term, the $\gamma N \Delta\pi$ Kroll-Ruderman term, and offers information on 
the $q$ dependence and the sign of the $s$ and $d- wave$ amplitudes of $N^\ast$(1520) 
$\rightarrow\Delta N$ decay. This information is a good test of quark models 
[14] which is passed by the "relativised" constituent quark models 
[15,16,17].
In the $\gamma p\rightarrow\pi^0\pi^0 p$ reaction the $\gamma N \rightarrow 
N^\ast(1520)\rightarrow\Delta\pi$ is shown to be an important mechanism by 
itself (there is no $\Delta$ Kroll-Ruderman term here), and it is perfectly visible in 
the experimental invariant mass distribution [5].

The ($\gamma$,$2\pi$) reaction has also relevance in nuclear physics. Inclusive 
cross section for the ($\gamma$,$\pi^+\pi^-$) have been calculated in [18] and 
the inclusive ($\gamma$,$\pi^0\pi^0$) reaction is measured in 
[19]. Calculations for ($\gamma$,$\pi^+\pi^-$) and the ($\gamma$,$\pi^0\pi^0$) 
coherent two pion production in nuclei have been performed in [20] and the 
cross sections are found to be very different in two different charge 
channels, with patterns in the energy and angular distributions linked to 
isospin conservation. Similarly, exchange currents for the ($\gamma$,$\pi^+$)
reaction are constructed from ($\gamma$,$\pi\pi$) when a pion is produced 
off-shell and absorbed by a second nucleon. These exchange currents give an 
important contribution to the ($\gamma$,$\pi^+$) cross section at large 
momentum transfer [21].
                                              
The discussions above have served to show the relevance of the 
($\gamma$,$2\pi$) reaction and its implications in different processes. The 
extension of this kind of work to virtual photons should complement the 
knowledge obtained through the ($\gamma$,$2\pi$) and the related reactions. 
The coupling of the photons to the resonances depends on $q^2$ and
the dependence
can be different for different resonances. Hence, the interference of different 
mechanisms pointed above will depend on $q^2$ and with a sufficiently large 
range of $q^2$, one can pin down the mechanism of ($\gamma$,$2\pi$) with real 
or virtual photons with more precision than just with real photons, which would 
help settle the differences between present theoretical models.

However, there are already interesting two pion electroproduction
experiments selecting $\Delta$ in the final state. The
reactions are, $e p\rightarrow e^\prime\pi^+\Delta^{++}$
and $e p\rightarrow e^\prime\pi^+\Delta^0$ [22].
 It is thus quite interesting to extend present models of $(\gamma,2\pi)$
to the realm of virtual phtons
and compare with existing data. In the present paper we do so, extending the model of ref.[2]
to deal with
the electroproduction process. This model is flexible enough and one can select
the diagrams which contain
$\Delta \pi$ in the final state in order to compare directly
with the measured cross sections.

The extension of the model requires three new ingredients: the introduction of the zeroth
component of the
photon coupling to resonances (calculations where done in [2] in the Coulomb gauge,
$\epsilon^0$, where the
zeroth component is not needed),  the implementation of the $q^2$
dependence of the amplitudes, which
will be discussed in forthcoming sessions, and the addition
of the explicit terms linked to the $S_{1/2}$ helicity amplitudes which
vanish for real photons.

Experiments on ($\gamma_v$,$2\pi$) are presently being done in the  Thomas 
Jefferson Laboratory [23], both for $N \Delta$ and $N\pi\pi$ production. 

The model presented here can serve to extract relevant information from coming data, which will
improve appreciably the precission of present experiments.

 There has been earlier work on the particular problem which we are discussing
 here. In \cite{bartl} an approach for $\Delta \pi$ electroproduction close to
 threshold was done with special emphasis in determining the axial transition
 form factors. The approach uses the current algebra formalism and some of the
 input needed is obtained from the same electroproduction data. The delta is
 treated as an elementary particle but the cross sections  are folded with a
 Breit Wigner distribution in order to take into account approximately the
 effects of the finite width of the $\Delta$. The effects of the $D_{13}$
 resonance are also discussed and introduced in an approximate way. The
 novelties here with respect to this work would be the use of a different
 formalism, since we are relying on the Feynman diagrammatic approach. On the
 other hand all information needed here, some of which was not available at the
 time of ref.\cite{bartl},
 is already known such that clear predictions can be made. There is another
 important difference with respect to \cite{bartl} which is that the $\Delta$ is
 not considered as a final state but decays explicitly into $\pi N$ in the
 Feynman diagrams considered. This means that the $\Delta$ is treated as a
 propagator and the sum over polarizations is done in the amplitudes not in the
 cross sections as in \cite{bartl}. This allows one to keep track of angular
 correlations between the two pions which are lost if the $\Delta$ is considered
 as a final state. It also allows to take into account the delta width in a
 natural way since it just comes as the imaginary part of the inverse of the 
 $\Delta$ propagator. It also allows one to keep track of interference of
 different pieces, in particular those between the $\Delta$ Kroll Ruderman term
 and the excitation of the $N^\ast(1520)$ resonance followed by 
 $\Delta \pi$ decay,
 which is one of the important findings in the present reaction.
 
  Our model bears more similarity to the work of \cite{berends} where also a
  diagrammatic approach is followed. They introduce the minimal set of terms
  which are gauge invariant as a block, including the important $\Delta$ Kroll
  Ruderman term. In our model we also include the four terms of \cite{berends}
  but we have in addition four more terms, in particular the excitation of 
  the $N^\ast(1520)$ resonance followed by $\Delta \pi$ decay, which has a strong
  interference with the  $\Delta$ Kroll Ruderman term. In \cite{berends} a
  formalism is used which respects Ward identities and leads to a gauge
  invariant amplitude in the presence of different electromagnetic form factors
  for the different terms of the model appearing for virtual photons. We have
  also followed this formalism in our approach. 
    
    For the construction of the currents for resonance excitation we follow
    closely the work of \cite{devendish} and take the convention of
    \cite{rollnik} for the definition of the helicity amplitudes. Altogether the
    formalism in the present paper diverts somewhat from the one used for real
    photons in [1,2] where many simplifications could be done, but in the
    case of real photons we regain the results of  [1,2], although the 
    use of new conventions forces the change of some sign. In order to avoid
    confusion the Lagrangians used and new conventions are now written in detail
    in a section.

\section{Model for $e N\rightarrow e^\prime\Delta\pi$.}

We will evaluate cross sections of virtual photons integrated over all the 
variables of the pions and the outgoing nucleon. In this case the formalism is 
identical to the one of inclusive $eN\rightarrow e^\prime X$
scattering [28,29]
or pion electroproduction after integrating over the pion variables 
[30,31]. The ($e$, $e^\prime$) cross section is given by 
\begin{eqnarray}
\frac{d\sigma}{{d\Omega^\prime}{dE^\prime}} &  = & \frac{{\alpha}^2}{q^4}
\frac{k^\prime}{k}\frac{-q^2}{1-\epsilon}\frac{1}{2\pi e^2}
[(W^{xx} + W^{yy})+ \\ \nonumber
& & 2\epsilon(\frac{-q^2}{\vec{q}\, ^2})W^{00}]
\end{eqnarray}
where $\alpha = e^2/{4\pi}$ is the fine structure constant, $e$ the electron 
charge, $q^\mu$ the momentum of the virtual photon and $k$, $k^\prime$
the momenta of the
initial and the final  electron and $\epsilon$ the polarization 
parameter of the photon, which is given by 
\begin{equation}
\epsilon = [1 - \frac{2\vec{q}\, ^2}{q^2}tg^2\frac{\theta_e}{2}]^{-1}
\end{equation}
with $\theta_e$  the angle of the scattered electron. All variables are given 
in the lab frame and the $z$ direction is taken along the direction of 
the virtual  photon, $\vec{q}$. Furthermore, the hadronic tensor is given here by
\begin{eqnarray}
W^{\mu\nu} & = & \int\frac{d^3p_2}{(2\pi)^3}\frac{M}{E_1}\frac{M}{E_2}
\int\frac{d^3p_4}
{(2\pi)^3}\frac{1}{2w_4}\int\frac{d^3p_5}
{(2\pi)^3}\frac{1}{2w_5} \\ \nonumber
& & \overline{\sum}\sum T^\mu T^{\nu\ast}(2\pi)^4\delta(q+p_1-p_2-p_4-p_5)
\end{eqnarray}
where $p_1$, $p_2$, $p_4$, $p_5$ are the momenta of the initial, final nucleon, and the two
pions and $T^\mu$ is the matrix element of the $\gamma_v 
N_1\rightarrow N_2\pi_4\pi_5$ process. Note that the phase space accounts for the decay of the $\Delta$
into $N \pi$ explicitly, hence the finite width of the $\Delta$ is automatically taken into account.
The terms contributing to $T^\mu$
are given below.

The expression of eq.(1) can be conveniently rewritten as [30]
\begin{equation}
\frac{d\sigma}{{d\Omega^\prime}{dE^\prime}} = \Gamma(\sigma^T_{\gamma_v} + 
\epsilon\sigma^L_{\gamma_v})=\Gamma\sigma_{\gamma_v}
\end{equation}
where $\sigma^T_{\gamma_v}$, $\sigma^L_{\gamma_v}$ 
are the transverse,longitudinal cross sections 
of the 
virtual photons and $\Gamma$ is given by
\begin{equation}
\Gamma = \frac{\alpha}{2\pi^2}\frac{1}{-q^2}\frac{k^\prime}{k}\frac{1}{1 - 
\epsilon}K_\gamma
\end{equation}
\begin{equation}
K_\gamma = \frac{s - M^2}{2M}\hspace{0.25cm}; s = (q^0 + M)^2 -\vec{q}\, ^2
\end{equation}

The corresponding cross sections $\sigma_{\gamma_v}^T$ and
$\sigma_{\gamma_v}^L$ are easily induced from eqs. (1) and (4). The
term with the combination $W^{xx}$ + $W^{yy}$ in eq. (1) gives rise
to the transverse cross section while the term proportional to
$W^{00}$ gives rise to the longitudinal one.

In the limit of the real photons, when $q^2\simeq0$, $K_\gamma$ is
the lab momentum
of the photon, and only the transverse cross section contributes, in which case 
$\sigma^T_{\gamma_v} = \sigma_\gamma$, the cross section of real photons.

For the model of the $\gamma_v N \rightarrow\Delta\pi$ reaction we take the 
same diagrammatic approach as in ref.[2] and select the diagrams which have a $\Delta$ in the 
final state. The diagrams which  contribute to the process are depicted in 
fig.1. The contribution of each one of the diagrams is readily evaluated from 
the Lagrangians written in appendix A1. The Feynman rules for the diagrams
 are collected in appendix A2. The coefficients, coupling constants and the
  form factors are collected in appendix A3. Finally, the amplitudes for each one
  of the terms are written in appendix A4 for each charge state.

Out of the 20 terms in [2] for the general $(\gamma,2\pi)$ reaction only 8 terms contain a $\Delta$
and a pion in the final state which are the terms collected in fig.1.

\begin{figure}[h]
\centerline{\protect
\hbox{
\psfig{file=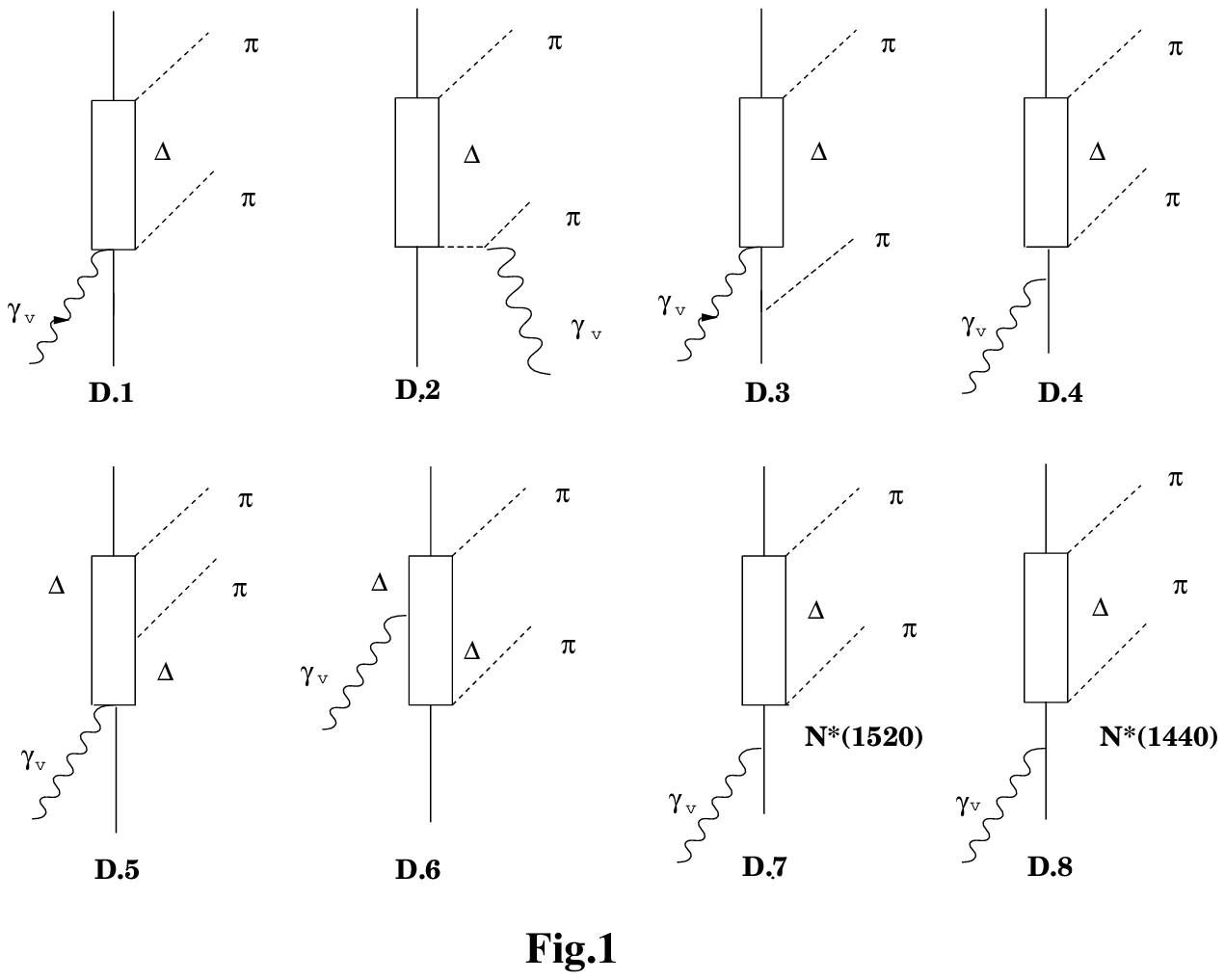,height=8cm,width=12.5cm,angle=0}}}
\caption{Feynman diagrams used in the model for $\gamma_v p\rightarrow
\pi\Delta$}
\end{figure}

\section{Electromagnetic transitions for Roper and N*(1520) resonances}

We follow the paper from Devenish et al. [26] in our approach to these transitions.
As we are working with virtual photons we need to care about these
couplings and hence include terms which vanish for real photons.
For the diagram D.8, which involves the Roper
excitation as depicted in the fig. 1, we can write the
corresponding electromagnetic current as :
\begin{equation}
J_{e.m.}^\mu= \bar{u}_{N^{\ast}}(p^\prime)
[\frac{\bar{F_2}(q^2) i\sigma^{\mu\nu}q_\nu}{m+m^\ast}
+\bar{F_3}(q^2)(q^\mu-\frac{q^2}{m^\ast-m}\gamma^\mu)]
{u}_N(p)
\end{equation}
where $\bar{F}_{2,3} (q^2)$ are the electromagnetic form factors for the 
$N-N^\ast$ transition (which already include the proton charge), 
$q^\mu=(q^0,\vec{q})$ is the momentum of the virtual photon and the
$m$, $m^\ast$ the masses of the nucleon and the $N^\ast(1440)$ respectively.
We can rewrite these form factors
in terms of $F_1$, $F_2$, defined as:
 $F_1=\frac{\bar{F}_3}{m^\ast-m}$ and $F_2=\frac{\bar{F}
_2}{m+m^\ast}$. The current of eq. (7) coincides with the
one in [26] substituting there: $G_1=-F_1$ and
$G_2=-\frac{2F_2}{m^\ast - m}$. We write our vertex functions 
as $V^{\mu}\epsilon_{\mu}$=$ -i
J^{\mu}\epsilon_{\mu}$. By keeping terms up to $(q/m)$ in a non relativistic
reduction of the matrix elements of the Dirac gamma matrices we find:

\begin{equation}
\hspace{-1.5cm} 
V^\mu_{\gamma N N^\ast} = \left\{\begin{array}{c} 
i\frac{\vec{q}\, ^2}{2m}F_2(q^2) - i\vec{q}\, ^2
(1+\frac{q^0}{2m})F_1(q^2)  \\ 
F_2(q^2)[i\vec{q}\,\frac{q^0}{2m}+(\vec{\sigma}\times\vec{q})
(1+\frac{q^0}{2m})]
-F_1(q^2)[i\vec{q}q^0(1+\frac{q^0}{2m})+q^2\frac{1}{2m}
(\vec{\sigma}\times\vec{q})]  
\end{array} \right\}
\end{equation}

Next we construct the helicity amplitudes for our transition. There are many
works where the helicity amplitudes are calculated
\cite{rollnik,alvarez,carlson2,carlson,stoler}.
In what follows we adjust to the formalism of ref. [27]. Then $A_{1/2}$ and
$S_{1/2}$ can be written as:
\begin{equation}
A_{1/2}^{N^\ast} =\sqrt{\frac{2\pi\alpha}{q_R}}\frac{1}{e}\langle
N^\ast,J_z=1/2|\epsilon_\mu^{(+)}\cdot J^{\mu}|N,S_z=-1/2\rangle
\end{equation}
\begin{equation}
S_{1/2}^{N^\ast} =\sqrt{\frac{2\pi\alpha}{q_R}}\frac{|\vec{q}\, |}{\sqrt{Q^2}}
\frac{1}{e}\langle N^\ast,J_z=1/2|\epsilon_{\mu}^{(0)}\cdot J^{\mu}|N,S_z=1/2
\rangle
\end{equation}
where $q_R$ is the energy of an equivalent real photon, $(W^2 - m^2)/(2W)$ and
$W$ is the photon-proton center of mass energy.

The tranverse polarization vectors are:
\begin{equation}
\epsilon_\mu^{(\pm)}=\frac{(0,\mp1,-i,0)}{\sqrt{2}}
\end{equation}
and $\epsilon_\mu^{0}$ = $\frac{1}{\sqrt{Q^2}}(q,0,0,q^0)$ 
normalized to unity, satisfying 
$\epsilon_\mu^{0}\cdot q^\mu$ =$g^{\mu\nu}\epsilon_\mu^{0}\cdot\epsilon_\nu^{(\pm)}=0$ with
$\vec{q}$ in $z$ direction and
$Q^2=-q^2$.

Using our electromagnetic current the helicity amplitudes are given by:
\begin{equation}
A_{1/2}^{N^\ast} =\sqrt{\frac{2\pi\alpha}{q_R}}\frac{1}{e} 
[F_2(q^2) \sqrt{2} q(1 +\frac{q^0}{2m})+ 
F_1(q^2) Q^2 \sqrt{2}\frac{q}{2m}]
\end{equation}
and   
\begin{equation}
S_{1/2}^{N^\ast} =\sqrt{\frac{2\pi\alpha}{q_R}}
\frac{\vec{q}\, ^2}{e} 
[-F_2(q^2) \frac{1}{2m} + F_1(q^2) (1 +\frac{q^0}{2m})]
\end{equation}

Inverting these equations we can get the electromagnetic form factors 
in terms of the helicity
amplitudes. The experimental helicity amplitudes $A^p_{1/2}$ and $S^p_{1/2}$
which we use are taken from \cite{lie} which uses data from \cite{genhart}.

In the case of the $N^\ast(1520)$ resonance we can take the same steps as
above. Following \cite{devendish} we write the relativistic current as:

\begin{equation}
J_{e.m.}^{\mu} = G_1(q^2)J_1^{\mu}+
G_2(q^2) J_2^{\mu}+
G_3(q^2) J_3^{\mu}
\end{equation}
with 
\begin{equation}
J_1^{\mu}=\bar{u}_\beta(q^\beta\gamma^\mu -q \!\!\!\!\, / g^{\beta\mu})u
\end{equation}
\begin{equation}
J_2^{\mu}=\bar{u}_\beta(q^\beta p^{\prime\mu} -p^\prime\cdot q
g^{\beta\mu})u
\end{equation}
\begin{equation}
J_3^{\mu}=\bar{u}_\beta(q^\beta q^\mu - q^2g^{\beta\mu})u
\end{equation}
and $G_1$,$G_2$,$G_3$ are the electromagnetic form factors for this vertex
and $p^\prime$ is the momenta of the resonance.

Taking a non relativistic reduction as done before and using 
$u_\mu$ Rarita-Schwinger spinors
in the c.m. of the resonance, the vertex takes an expression given by:
\\

Scalar part:
\begin{equation}
V^0_{\gamma N N^{\prime\ast}}=i(G_1+G_2 p^{\prime 0} + G_3 q^0)\vec{S}^\dagger
\cdot\vec{q}
\end{equation}
and the vector part:
 
\begin{eqnarray}
V^i_{\gamma N N^{\prime\ast}}=-i
[(\frac{G_1}{2m} - G_3)(\vec{S}^\dagger\cdot\vec{q})\, \vec{q} - 
\\ \nonumber & & 
\hspace{-5cm} 
iG_1\frac{\vec{S}^\dagger\cdot\vec{q}}{2m}(\vec{\sigma}\times\vec{q})
-\vec{S}^\dagger \{G_1(q^0+\frac{\vec{q}\, ^2}{2m}) 
+ G_2 p^{\prime 0} q^0 + G_3
q^2\}]
\end{eqnarray}

Using again eqs.(9) and (10) we calculate the $A_{1/2}$ and $S_{1/2}$
helicity amplitudes for the $N^\ast(1520)$. In addition, in this case
we also have the $A_{3/2}$ helicity amplitude which is given in 
\cite{rollnik} by:

\begin{equation}
A_{3/2}^{N^\ast} =\sqrt{\frac{2\pi\alpha}{q_R}}\frac{1}{e}\langle
N^\ast,J_z=3/2|\epsilon_\mu^{(+)}\cdot J^{\mu}|N,S_z=1/2\rangle
\end{equation}

The expressions for the helicity amplitudes  obtained using
the current in eqs.(14-17) are:
\begin{equation}
A_{3/2}^{N^{\prime\ast}}=\sqrt{\frac{2\pi\alpha}{q_R}}\frac{1}{e} 
[{G_1(q^0 +\frac{\vec{q}\, ^2}{2m}) + G_2 q^0 p^{\prime 0} +G_3 q^2}]
\end{equation}
\begin{equation}
A_{1/2}^{N^{\prime\ast}}=\sqrt{\frac{2\pi\alpha}{q_R}}\frac{1}{e}\frac{1}{\sqrt{3}} 
[{G_1(q^0 -\frac{\vec{q}\, ^2}{2m}) + G_2 q^0 p^{\prime 0} +G_3 q^2}]
\end{equation}
\begin{equation}
S_{1/2}^{N^{\prime\ast}}=-\sqrt{\frac{2\pi\alpha}{q_R}}\frac{1}{e}\sqrt{\frac{2}
{3}} q
[{G_1 + G_2 p^{\prime 0} +G_3 q^0}]
\end{equation}

From these three equations we can get the $G_1,G_2,G_3$ form factors in terms of
the helicity amplitudes. The data for $S_{1/2}$ are taken from
\cite{rollnik} and for $A_{1/2}$ and $A_{3/2}$ from \cite{rollnik,burcketpri}
\footnote{We note that the formalism followed here for the vertices
of the $N^\ast(1520)$ is different from that of [2], but the same results
are obtained for real photons.}.

The other important vertex in our model corresponds to the
$\Delta - N$ electromagnetic transition. As discussed in  
\cite{alvarez,carlson2}, the most
important transition is the magnetic dipole ($M_{1^+}$) transition
while the electric quadrupole ($E_{1^+}$) and scalar quadrupole ($S_{1^+}$)
transitions are small at momentum transfers between the photon point
to $Q^2$=1.3 GeV$^2$. The values given in \cite{devendish} for the
ratio $E_{1^+}/M_{1^+}$ ($S_{1^+}/M_{1^+}$) are -0.02 to 0.02 (-0.025 to -0.06) for
 $Q^2$=0 to 1.3 GeV$^2$ respectively. We take the $\gamma N\Delta$
 transition current from [39] where the same non relativistic expansion
 done here, keeping terms of order $O(p/m)$ for the Dirac matrix elements, is
 done. A good reproduction of the data for electroproduction of one pion
 was obtained there in a wide range of energies around the $\Delta$ resonance
 and different $Q^2$. The vertex for this electroproduction transition is
 given by eq. (40) in appendix A2 and the form factor used is given by eq. (61)
 in appendix A3.

\section{Gauge invariance and form factors}
 
Gauge invariance is one of the important elements 
in a model involving photons and implies that
\begin{equation}
T^\mu q_\mu = 0
\end{equation}

The explicit expressions for $T^\mu$, keeping the four components, as given 
in appendix A4, allow one
to check explicitly the gauge invariance. The block of diagrams D1, D2, D4 and D6 form together a gauge 
invariant set.  The rest of the diagrams in which the photon directly excites 
a resonance from a nucleon are gauge invariant by themselves. However, some caution must be observed when 
imposing eq. (24). Indeed, in diagram D2 the intermediate pion is off-shell and  induces a
strong $\pi N \Delta$ transition form factor, $F_\pi(p^2)$, for which we take a usual monopole form factor
(see eq.(53) appendix A3). The constraints of eq.(24) forces this form factor to appear in the other terms
of the block of diagrams which are gauge invariant. However, as discussed in the study of the $e 
N\rightarrow e^\prime N\pi$ reaction in [39], and as can be easily seen by inspection of the diagrams
and the amplitudes, the constraint of eq. (24) still requires
the equality of four electromagnetic form factors,

\begin{equation}  
F_1^p(q^2) = F_1^\Delta(q^2)= F_{\gamma\pi\pi} = F_c(q^2)
\end{equation}

The form factors of eq. (25), if the strict Feynmann rules of the
appendix A4 are followed, are respectively
 the $\gamma N N$, $\gamma\Delta\Delta$, $\gamma\pi\pi$ and 
 $\gamma\Delta N\pi$ ones.
 These form factors are usually parametrized in different forms, as seen in appendix A3, except for
$F_1^p (q^2)$ and $F_1^\Delta (q^2)$ which are taken equal, 
as it would come from
ordinary quark models.

Although the model is gauge invariant with the prescription
of eq. (25) there is the inconvenience that the results
depend upon which one of the three form factors we take for all of them.

In the next section we discuss the uncertainties which come from this arbitrary choice of form 
factor. We should note however, that the dominant term, by large, is the 
$\Delta$ Kroll Ruderman 
and pion pole terms. This is also so in the test of gauge invariance 
where the two terms involving 
the $F_1^p(q^2)$ form factor in diagrams D4, D6 give only recoil 
contributions of the order of 
O($p_\pi$/m) in eq. (24). This justifies the use of $F_c(q^2)$ or $F_{\gamma\pi\pi}(q^2)$ for all the 
form factors.

There is, however, another way to respect gauge invariance, while at the same
time using different form factors which is proposed in \cite{berends} and 
to which we refer in what follows as Berends et al. approach. 
In the next section  we will show the results
from both approaches and will discuss them. 

There are many papers in the
literature following the Berends et al. approach in order to explain the experimental
results with this gauge invariant set of diagrams \cite{kiel,driver,mont,bartl2}. 

Another of the common approaches to this problem in the past has been
the use of current algebra \cite{joos,adler,carro}. In this case, close
to threshold of $\Delta\pi$ production, the axial
vector current $\langle\Delta| A_\mu |N\rangle$ is the dominant term.
 In the soft pion limit this axial vector current is shown to be
 equivalent \cite{adler} to the one of the nucleon. The $\pi\Delta$
 electroproduction data are hence used in those works to determine the
 nucleon axial form factor.
 
 Our model is more general since one obtains explicit $p_\pi^\mu$ dependence, 
 for instance from the pion pole term (diagram D.2 fig. 1). The
 dominance of the $\Delta N\pi\gamma$ Kroll Ruderman at threshold
 (independent of $p_\pi^\mu$) offers however some support to the
 assumptions made in those works, as quoted in [30]. The diagrammatic
 approach relies however on the explicit use of the four form factors
 of eq. (25). We shall see in the next section the influence of the
 different terms close to threshold.
 
In \cite{bartl3}, following the diagrammatic approach with the minimal
set of gauge invariant terms, fits to the data are carried fixing 
$F_{\gamma\pi\pi}$ from $\rho$ vector meson dominance and setting the other
three form factors equal in order to determine this latter common form factor.
A qualitative agreement with the $F_1^p(q^2)$ form factor obtained from others
sources is found.

With the gauge invariant prescription of \cite{berends} using different
form factors, different fits to the data are carried fixing some of the
form factors and determining others. We shall follow this latter approach
, but with our enlarged set of Feynman diagrams 
 including
the explicit decay of the $\Delta$ into $\pi N$, which allows one to keep track
of correlations between the pions if wished. We will show the sensitivity
of the results to different form factors and the weight of the different
terms as a function of energy and $Q^2$.

The explicit formulas of the gauge invariant set in the presence of the
different form factors are taken from eq. (17) of ref. \cite{berends}
and, after
the non relativistic reduction is done, the expressions used here 
are shown in appendix A5.

\section{Results and discussion}

We have tested our results with the experimental data of refs. [22,30]. We show
the cross section of $\gamma_v p\rightarrow\Delta^{++}\pi^-$ and
$\gamma_v p\rightarrow\Delta^0\pi^+$ ($\Delta^0\rightarrow\pi^- p$), as a function
of W, the virtual photon-proton ($\gamma_v p$) center of mass energy, and for
different values of $Q^2$. We have

\begin{equation}
W^2 = -Q^2 + M^2 + 2M\nu\,  ;\nu = E - E^{\prime}
\end{equation}
with
\begin{equation}
Q^2=-q^2=4EE^{\prime}sin^2(\frac{\theta_e}{2})
\end{equation}

 We compare in fig. 2 the results for delta
photoproduction (real photons) with the experimental data \cite{ABBA} and
we see that the agreement found is quite satisfactory.

\begin{figure}[h]
\centerline{\protect
\hbox{
\psfig{file=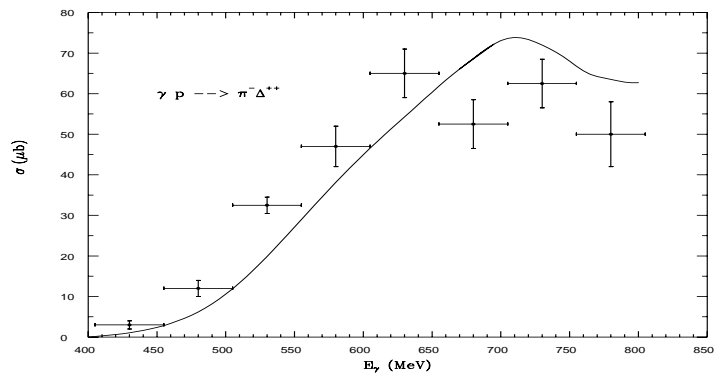,height=5cm,width=9.5cm,angle=-90}}}
\caption{\small{Cross section for the $\gamma p\rightarrow\Delta^{++}\pi^-$ reaction. Experimental
data from \cite{ABBA}.}}
\end{figure}

In order to compare our results with experiment for virtual photons we 
calculate first the cross sections at $Q^2$ = 0.6 $GeV^2$.

 Figure 3 contains three different calculations. Two of them correspond
to using all form factors equal (which we set to $F_{\gamma\pi\pi}$)
with two different values of $\lambda_\pi^2$, 0.5 $GeV^2$ and 0.6 $GeV^2$. 
We see that the cross section increases by about 10 $\%$ when going from
$\lambda_\pi^2$=0.5 $GeV^2$ and $\lambda_\pi^2$=0.6 $GeV^2$. We also
show the results taking $F_1^p$, $F_1^\Delta$ with their values
given in appendix A3 and setting $F_c=F_{\gamma\pi\pi}$ with $\lambda_\pi^2$=0.6
 $GeV^2$. This latter calculation is not gauge invariant. However we see
that the deviation with respect to the gauge invariant one assuming all form
factors equal is very small. This reflects the fact that the relevant terms
in the model are those involving $F_{\gamma\pi\pi}$ and $F_c$, the pion pole and
$\Delta$ Kroll Ruderman terms.

\begin{figure}[h]
\centerline{\protect
\hbox{
\psfig{file=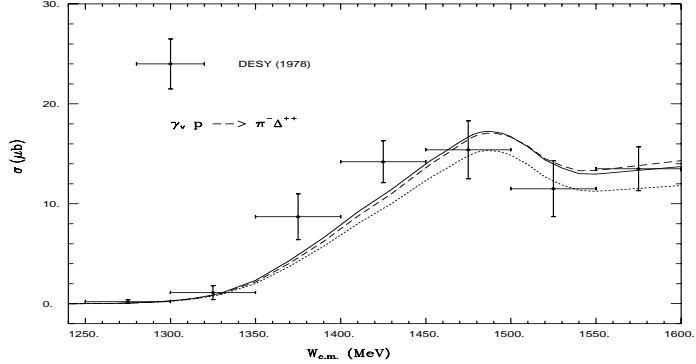,height=5cm,width=9.5cm,angle=-90}}}
\caption{\small{Cross section for $\gamma_v p\rightarrow\pi^-\Delta^{++}$ 
as a function
of the $\gamma_v p$ center of mass energy. The curves correspond to $Q^2$ = 0.6 
$GeV^2$. Dashed line: $F_1^p,F_1^\Delta$,
from appendix A3 and
$F_c$=$F_{\gamma\pi\pi}$ with $\lambda_\pi^2$ = 0.6 $GeV^2$. 
Continuous line: $F_1^p=
F_1^\Delta=F_c=F_{\gamma\pi\pi}$ with  $\lambda_\pi^2$ = 0.6 $GeV^2$.
Dotted line: same as continuous line with $\lambda_\pi^2$ = 0.5 $GeV^2$}}
\end{figure}

\begin{figure}[h]
\centerline{\protect
\hbox{
\psfig{file=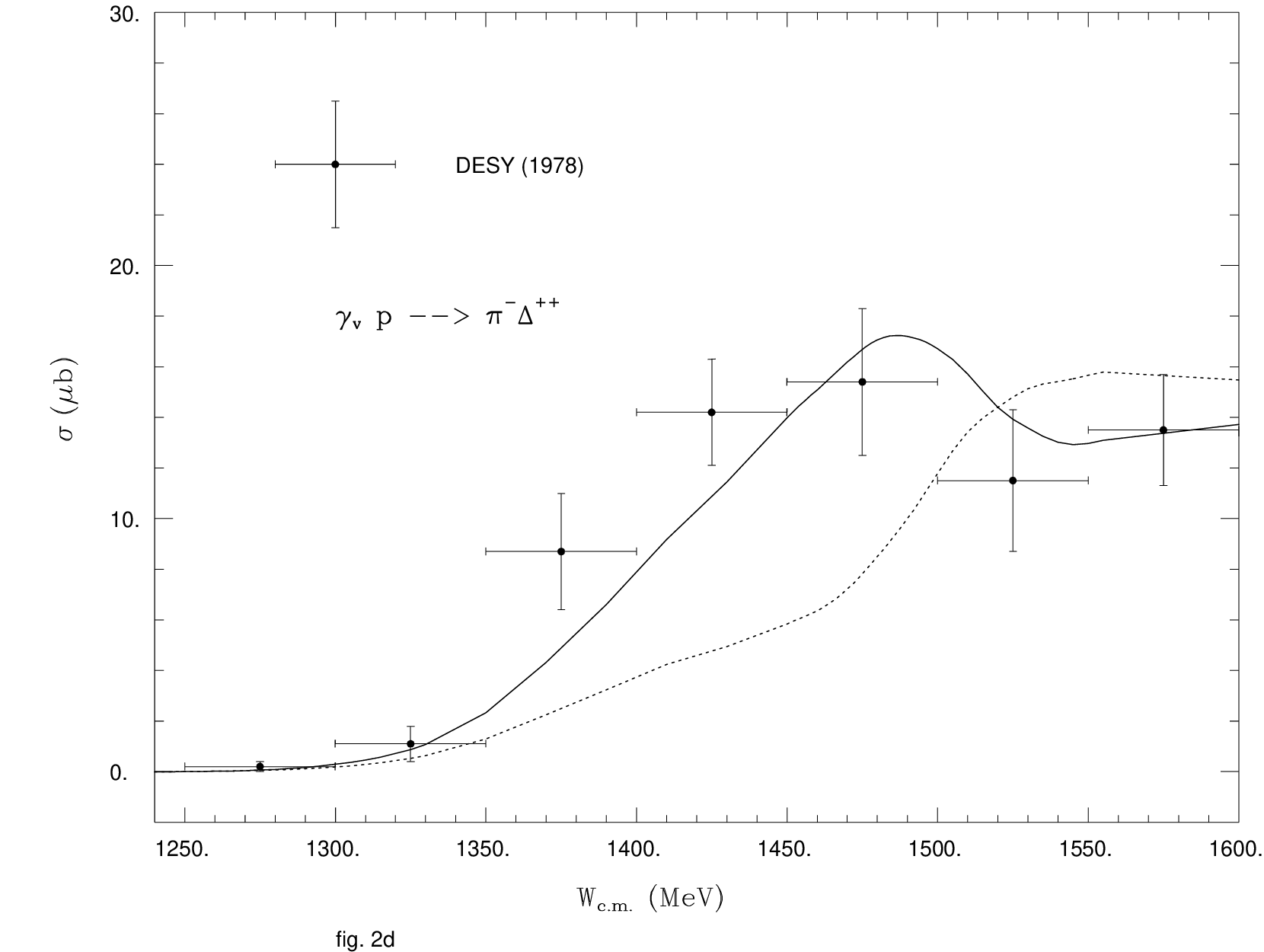,height=5cm,width=9.5cm,angle=-90}}}
\caption{\small{Continuous line: Cross section 
for $\gamma_v p\rightarrow\pi^-\Delta^{++}$ at $Q^2$= 0.6 $GeV^2$ using
$F_1^p=F_1^\Delta=F_{\gamma\pi\pi}=F_c$ with $\lambda_\pi^2$=0.6 $GeV^2$
and $\tilde{f}_{N^{\ast\prime}\Delta\pi}=-0.911$,  
$\tilde{g}_{N^{\ast\prime}\Delta\pi}$=0.552. Dotted line: same as continuous line
but with $\tilde{f}_{N^{\ast\prime}\Delta\pi}$=0.911 and 
$\tilde{g}_{N^{\ast\prime}\Delta\pi}=-0.552$. See appendices A1 and A2 for 
the vertices and coupling constants used.}}
\end{figure}

\begin{figure}[h]
\centerline{\protect
\hbox{
\psfig{file=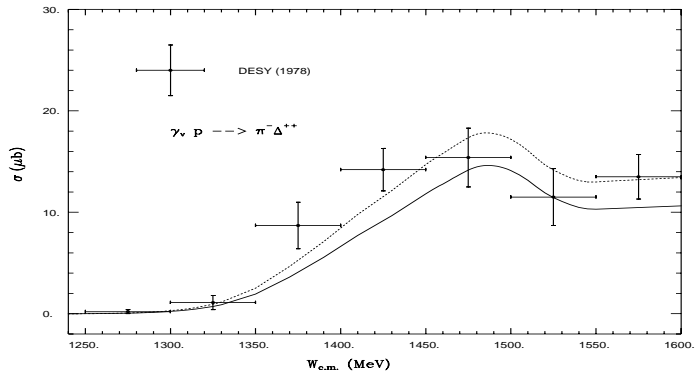,height=5cm,width=9.5cm,angle=-90}}}
\caption{\small{Cross sections from $\gamma_v p\rightarrow\Delta^{++}\pi^-$ with
Berends formalism included.
Continuous line: $F_1^p, F_1^\Delta, F_{\gamma\pi\pi}=F_c$ with
$\lambda_\pi^2= 0.5$ $GeV^2$. Dotted line: same as continuous line with the
parameter for $F_{\gamma\pi\pi}$ 0.5 $GeV^2$ and for $F_c$ is 0.8 $GeV^2$.}}
\end{figure}

In refs. [2,14], two solutions for the
coupling of the $N^\ast(1520)$ to the $\Delta$ in $s$ and $d- waves$, 
differring only in a global sign,
 were found
from the respective decay widths. Only a sign was compatible with
the experimental $(\gamma,\pi\pi)$ data, because of the strong interference
between the $\gamma N\rightarrow N^\ast(1520)\rightarrow\Delta\pi$ term
and the Kroll Rudermam one. Here, in the new formalism the amplitude
of the $\gamma N\rightarrow N^\ast(1520)$ transition changes sign
and consequently the signs of the former $N^\ast(1520)\rightarrow\Delta\pi$
couplings must be changed. In fig. 4 we show the results of
$\gamma_v p\rightarrow\pi^-\Delta^{++}$ at $Q^2$=0.6 $GeV^2$ using the two different
signs for these couplings. We can see that the data favour clearly one of the
signs. This was also observed in the $\gamma p\rightarrow\pi^+\pi^- p$
reaction in [14]. Fig. 4 shows the magnitude of the interference between
those terms. Neglecting the $\gamma
p\rightarrow N^\ast(1520)\rightarrow\Delta\pi$ process leads
to results in between the two curves shown in the figure where one does not
see any peak. The peak shown by the data around W=1480 MeV is hence
not reminiscent of a $\Delta$ but a peak coming from the constructive
interference of the terms mentioned above.

Now we evaluate the cross section using Berends gauge invariant approach
with different form factors \cite{berends} using the formulae of appendix A5 for
the set of the four gauge invariant terms. We show the results in fig. 5.
The continuous line in the figure
is obtained with this prescription using the form factors of appendix
A3 for the $F_1^p$, $F_1^\Delta$ but setting $F_c=F_{\gamma\pi\pi}$ with
$\lambda_\pi^2$ = 0.5 $GeV^2$.

We can see that these results are remarkably similar to those of the
dotted line in fig. 3 where $F_c$ and $F_{\gamma\pi\pi}$ had the same
values as here but $F_1^p$, $F_1^\Delta$ were set equal to
$F_{\gamma\pi\pi}$ in order to preserve gauge invariance.

Once again we can see that the terms involving $F_1^p$ and $F_1^\Delta$
are relatively unimportant, or in any case that setting these form factors
equal to $F_{\gamma\pi\pi}$ provide a gauge invariant result very close
to the one obtained with the more general Berends prescription of [25].

The dotted line in fig. 5 corresponds to the same parametrization for
$F_c$ as for $F_{\gamma\pi\pi}$ but parameter $\lambda_c^2$= 0.8 $GeV^2$.
This shows the sensitivity of the results to $F_c$  which appears in the 
dominant Kroll-Ruderman term.
 
Given the dominance of the Kroll Ruderman term close to threshold production
of $\Delta\pi$
in our model and the dominance of the terms involving the axial form factor
in the current algebra approaches, we now show in fig. 6 the 
results taking
for $F_c$ the parametrization of the axial vector form factor (see appendix
A3), varying the value of $M_A$. The Berends formalism with different form
factors is used, and $F_1^p$, $F_1^\Delta$, $F_{\gamma\pi\pi}$ are taken
as in appendix A3. We can see that the experimental data favour values
of $M_A \simeq$ 1.16-1.23 $GeV$ very similar to the values determined in 
[30] (1.16$\pm$ 0.03) GeV or [44] (1.18$\pm$ 0.07) GeV.
\begin{figure}[h]
\centerline{\protect
\hbox{
\psfig{file=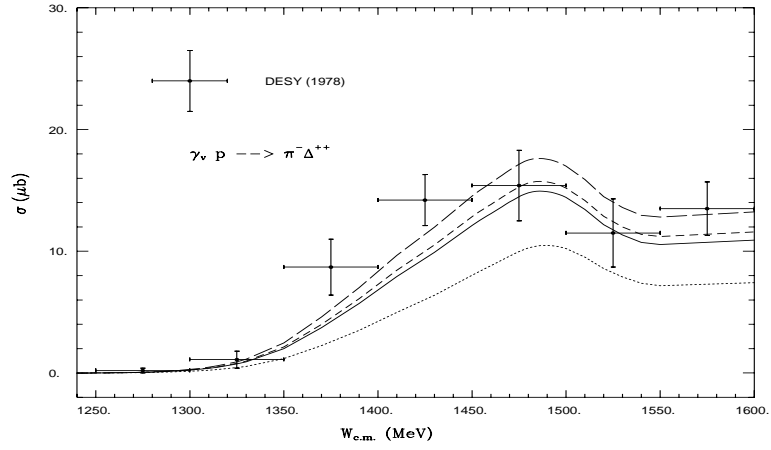,height=6cm,width=10.5cm,angle=-90}}}
\caption{\small{Cross sections for $\gamma_v p\rightarrow\pi^-\Delta^{++}$ within
Berends formalism. We use different form factors 
for $F_1^p$, $F_1^\Delta$, $F_{\gamma\pi\pi}$ from appendix A3.  
We take for the  
axial form factor, $F_A$, form the contact term. The parameter $M_A$ is changed from up
to down :1.32, 1.23, 1.16, 0.89 in GeV.}}
\end{figure}
\begin{figure}[h]
\centerline{\protect
\hbox{
\psfig{file=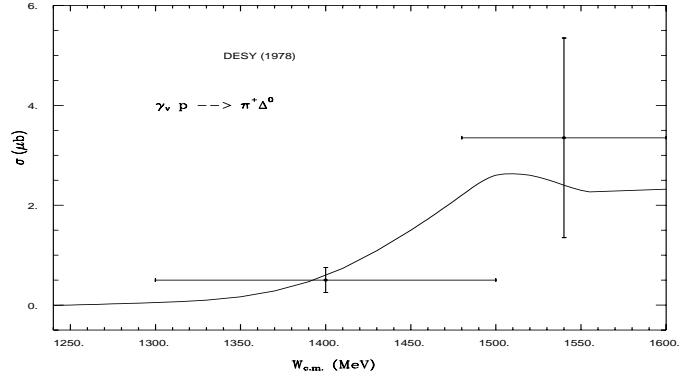,height=5cm,width=9.5cm,angle=-90}}}
\caption{\small{Cross section for $\gamma_v p\rightarrow\pi^+\Delta^0$ within Berends
formalism. We use $F_1^p, F_1^\Delta, F_{\gamma\pi\pi}$ from appendix
A3 and $F_A$ with $M_A$=1.16 GeV.}}
\end{figure}

In fig. 7 we show the cross section for $\gamma_v p\rightarrow\Delta^0\pi^+$
with Berends scheme and with a value of the axial form factor parameter of
$M_A =1.16$ GeV.  We find a good agreement with the scarce experimental data
but more data would be required to further check this channel.

In all these sets of figures for $\Delta^{++}$ electroproduction we find a good agreement with the experimental
results around the peak region
 1460 MeV$\le$ W $\le$ 1600 MeV in c.m. energy
at $Q^2$=0.6 $GeV^2$. As commented above, this peak comes from the 
interference between the $\Delta$ Kroll Ruderman and the $N^\ast(1520)$ 
excitation
terms, which was also found for real photons in [1,2,14]. A priori this interference
pattern could change for virtual photons since these two terms are affected
by different form factors, $F_c(q^2)$ for the contact term and the form factors
$G_1(q^2), G_2(q^2), G_3(q^2)$ that appear for the $\gamma N\rightarrow N^\ast(1520)$ transition.
 In practice we see that the interference survives in the case of virtual
 photons.

The $N^\ast(1520)$ excitation has a longitudinal coupling and so has the
$\Delta$. The inclusion of this longitudinal contribution could also 
blur the interference found with these
two terms for real photons. However, the results shown for virtual photons
show that the interference remains also in this latter case. We have checked
that if we go up to higher $Q^2$ the
interference still exists but the shape becomes flatter.

\begin{figure}[h]
\centerline{\protect
\hbox{
\psfig{file=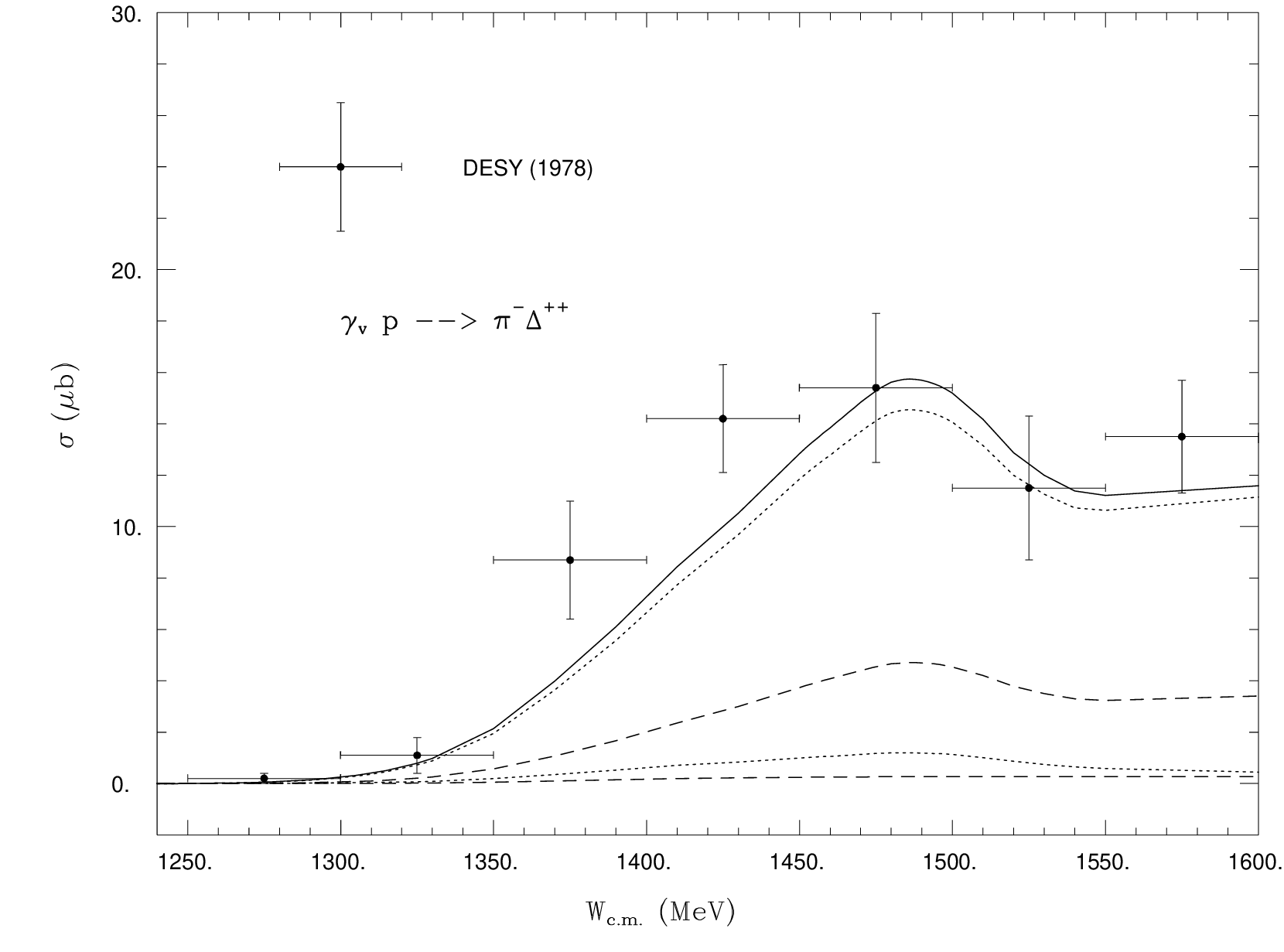,height=6cm,width=10.5cm,angle=-90}}}
\caption{\small{Continuous line: cross section for $\gamma_v p\rightarrow
\Delta^{++}\pi^-$ within Berends formalism and with the parameter 
$M_A$=1.23 GeV at $Q^2$=0.6 $GeV^2$. Dotted lines: From up to down we show the tranverse $\sigma_T$
and longitudinal $\epsilon\sigma_L$ cross section at $Q^2$=0.6 $GeV^2$ 
respectively. 
Short-dashed lines: From up to down the same as in dotted line for $Q^2$=1.2
$GeV^2$.}}
\end{figure}
\hspace{3cm}
\begin{figure}[h]
\centerline{\protect
\hbox{
\psfig{file=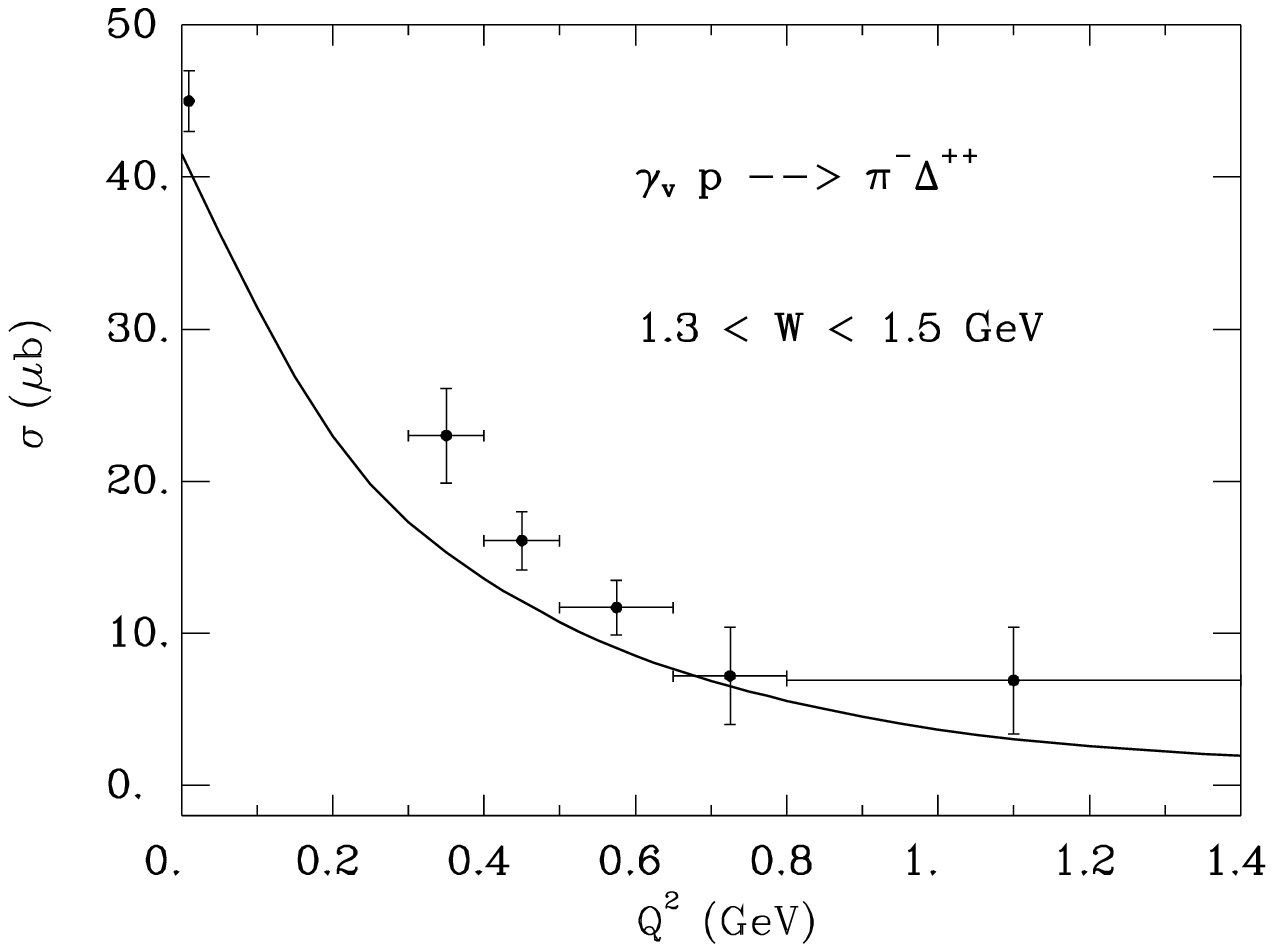,height=6cm,width=10.5cm,angle=-90}}}
\caption{\small{Cross section $\gamma_v p\rightarrow\Delta^{++}\pi^-$
as a function of the $Q^2$.  }}
\end{figure}

\begin{figure}[h]
\centerline{\protect
\hbox{
\psfig{file=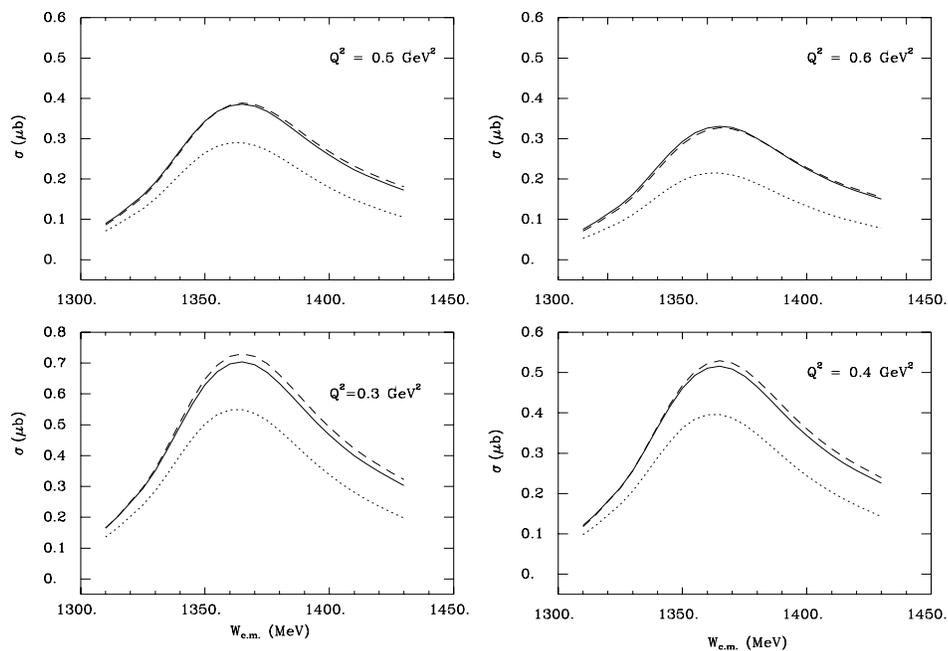,height=9cm,width=12.5cm,angle=-90}}}
\caption{\small{Continuous line: Total cross section from 
$\gamma_v\rightarrow\pi^-\Delta^{++}$ within Berens
formalism restricting $m_\pi\le \omega_{\pi^-}\le m_\pi$ + 10 MeV.
We use $F_1^p, F_1^\Delta$ from appendix A3 and $F_{\gamma\pi\pi}$ with 
$\lambda_\pi^2$= 0.5 $GeV^2$. Dotted lined:
Contribution of $\Delta$ Kroll Ruderman term using $F_A$ with $M_A$=1.16 GeV.
Short-dashed line: Contribution of $\Delta$ Kroll Ruderman term plus
the interference with the $N^\ast(1520)$ term.}}.
\end{figure} 
 It is interesting to show separate contributions for the longitudinal and
 transverse cross sections which are likely to be also measured at TJNAF.
 We find in fig. 8 that the shape of the curves for the transverse contribution is the
 same as for real photons and for the sum of the longitudinal and
 transverse cross sections for the virtual ones. We observe that the
 longitudinal contribution gives only
 a small background. This pattern appears for any intermediate values of $Q^2$
 between those shown in the figure.

 In fig. 9 we show a cross section as a function of the momentum transfer
 $Q^2$ and we compare with the experimental data from [22,30].
 
  We have made an average of the cross sections between 1300 to 1500 MeV c.m. energy in order
 to compare with the experiment. We observe that the trend of the data is well
 reproduced, but the absolute value is a little lower reflecting the
 discrepances with the data in fig. 8 in that range of energies.
 
 In fig. 10 we show some results which can be a guideline for an experimental
 analysis. We concentrate on a hypothetical measurement that strengtens
 the contribution of the Kroll Ruderman term in order to optimize the chances
 to obtain an accurate value for the contact form factor $F_c(q^2)$. This
 can be accomplished by fixing the energy of the $\pi^-$ and hence putting
 the $\Delta^{++}$ on shell in the diagram D.1 of fig. 1. In fig. 10 we take 
  $m_\pi\le \omega_{\pi^-}\le m_\pi$ + 10 MeV and show the results obtained
 for different $Q^2$ selecting the Kroll Ruderman term alone, this term plus
 the interference with the $N^\ast(1520)$ and the total contribution. We see
 indeed that this magnitude is largely dominated by the Kroll Ruderman term.
  However the interference with the $N^\ast(1520)$ term is always present. Since
  the weight of the $N^\ast(1520)$ propagator is smaller at lower energies, we
  can also see that the region to the left of the $\Delta$ peak which appears
  in the figure is more suited to pin down the contribution of the Kroll
  Ruderman term. Yet, even at this lower energies the contribution of the
  interference term is still of the order of 20 per cent. This means that the
  accurate evaluation of $F_c(q^2)$ at level better than 20 per cent requires a
  careful analysis in which the interference term is explicitly considered.

\section{Conclusions}

We have calculated cross sections for the $\gamma_v p\rightarrow\Delta^{++}
\pi^-$ and $\gamma_v p\rightarrow\Delta^0\pi^+$ reactions, extending the model
of ref. [2] to virtual photons and selecting the diagrams which have a $\Delta$
in the final state.

The present calculations and comparison with the scarce experimental data are
sufficient to establish the fairness of the present model to deal with the
$\Delta\pi$ production process. In summary we could remark the following points:

Even if the data are scarce the agreement with them is good
up to W$\simeq$ 1.6 GeV and $Q^2\simeq$ 1.4 $GeV^2$ for the $\Delta^{++}$ channel.
However, it would be desirable to have data for different values of $Q^2$. In
the future such experiments are bound to be made in Thomas Jefferson Laboratory
and other experimental facilities. Also other channels should be measured as
well as total cross sections for $\gamma_v N\rightarrow\pi\pi N$ where the $\pi
N$ are not in a $\Delta$ state.

We have also shown that the peak in the cross section is due to an interference
between the $\Delta$ Kroll Ruderman term and the $N^\ast(1520)$ excitation
process followed by $\Delta\pi$ decay. This interference appeared in real
photons and is not destroyed for virtual ones in spite of the fact that
the 
electromagnetic form factors of the respective mechanisms are not exactly the
same. In addition, the contribution from the longitudinal couplings in the terms
involved in the interference does not destroy this effect. The 
experiments show clearly that interference of the two mechanisms is
contructive below the $N^\ast(1520)$ pole, which is consistent with the findings
in real photons and predictions of the relativized quark models.

Different sets of form factors have been used in our model in order to show the
sensitivity of the results to these changes. These tests should
be useful in view of the coming data and the possibility to extract relevant 
information from them.

Some plots can be useful for the experimental task. We have shown the 
separation
of the transverse and longitudinal cross sections and found that the
transverse one largely dominates the cross sections.

We have also shown a method aimed at obtaining the form factor for the Kroll
Ruderman term selecting a kinematics which maximizes its importance. Even then
we saw that an accurate extraction of this form factor requires the explicit
inclusion of the $N^\ast(1520)$ term in the analysis because of the important
interference of this term with the Kroll Ruderman one.

Finally, it is also interesting to note that the present model is just part of a more general 
$\gamma_v N \rightarrow\pi\pi N$ model which selects only the terms 
where a $\pi N$ pair of the final state appears
forming a $\Delta$ state. Both experiments and theoretical calculations on the different ($\gamma_v,\pi\pi)$ channels should be
encouraged.

\vspace{3cm}

{\bf Acknowledgements.}
We would like to thank J.A. G\'omez-Tejedor, F. Cano and L. Alvarez-Ruso
for useful discussions. We would like to acknowledge partial financial support for DGICYT contract 
number PB96-0753. One of us, J.C.N. would like to acknowledge the support
from the Ministerio de Educacion y Cultura.
\newpage

\begin{center}
APPENDIX
\end{center}

{\Large{\bf A1. Lagrangians.}}

\begin{equation}
 L_{\pi N N} = - \frac{f}{\mu}\overline{\Psi}\gamma^\mu\gamma_5
\partial_\mu\vec{\phi}\cdot\vec{\tau}\Psi
\end{equation}
\begin{equation}
 L_{\Delta \pi N} = - \frac{f^\ast}{\mu}\Psi^\dagger_\Delta S^\dagger_i(
\partial_i\phi^\lambda)T^{\lambda\dagger}\Psi_N  +  h.c.
\end{equation}
%\begin{equation}
% L_{\Delta \pi \gamma N} = - i q_{\pi}\frac{f^\ast}{\mu}\Psi^\dagger_\Delta 
%S^\dagger_iA_i\phi^\lambda T^{\lambda\dagger}\Psi_N  +  h.c.
%\end{equation}
\begin{equation}
 L_{\Delta \Delta \pi} = - \frac{f_\Delta}{\mu}\Psi^\dagger_\Delta 
S_{\Delta i}(\partial_i\phi^\lambda)T^\lambda_\Delta\Psi_\Delta  +  h.c.
\end{equation}
\begin{equation}
 L_{N^{\ast} \Delta \pi} = - \frac{g_{{N^\ast}\Delta\pi}}{\mu}
\Psi^\dagger_\Delta  
S^\dagger_i(\partial_i\phi^\lambda)T^{\lambda\dagger}\Psi_{N^\ast}  +  h.c.
\end{equation}
\begin{equation}
 L_{N^{\ast\prime} \Delta \pi} = i \overline{\Psi}_{N^{\ast\prime}}(
\tilde{f}_{N^{\ast\prime} \Delta \pi}-\frac{\tilde{g}_{N^{\ast\prime} 
\Delta \pi}}
{\mu^2}S_i^\dagger\partial_i S_j\partial_j)\phi^\lambda T^{\lambda\dagger}
\Psi_\Delta  +  h.c.
\end{equation}
\begin{equation}
 L_{N N \gamma} = - e\overline{\Psi}_N(\gamma^\mu A_\mu -\frac{\chi_N}{2m}
\sigma^{\mu\nu}\partial_\nu A_\mu)\Psi_N
\end{equation}
%\begin{equation}
% L_{\Delta N \gamma} = - \frac{f_{\Delta N \gamma}}{\mu}\Psi^\dagger_
%\Delta \epsilon_{ijk} S^\dagger_i(\partial_j A_k)T_3^\dagger\Psi_N  +  h.c.
%\end{equation}
%\begin{equation}
% L_{N^\ast N \gamma} = \frac{\tilde{f^N_\gamma}}{\mu}\overline{\Psi}_N
%\sigma^{\mu\nu}\partial_\nu A_  +  h.c.
%\end{equation}
%\begin{equation}
% L_{N^\ast N \gamma} = \overline{\Psi}_{N^\ast}
%\bar{F_3}(\frac{q^2}{m^\ast-m}\gamma^\nu - q^\mu) - 
%\frac{\bar{F_2}}{m+m^\ast}
%\sigma^{\mu\nu}\partial_\nu A_\mu\Psi_{N}
%\end{equation}
%\begin{equation}
% L_{N^{\ast\prime} N \gamma} = 
%\end{equation}
\begin{equation}
 L_{\pi\pi\gamma} = ie(\phi_+\partial^\mu\phi_- - \phi_-\partial^\mu\phi_+)A_\mu
\end{equation}

Instead of writing the explicit expressions for the terms involving the 
photon and the
excitation of resonances like $L_{\Delta N \gamma}$, $L_{N^\ast N \gamma}$,
 $L_{\Delta\pi\gamma N}$ and $L_{N^{\ast\prime} N \gamma}$, we
 address the reader directly to eqs. (40, 43-44, 50, 51-52) respectively which provide the vertex function
 ($L\rightarrow - V^\mu \epsilon_\mu$).

In the former expressions $\vec{\phi}$, $\Psi$, $\Psi_\Delta$, $\Psi_{N^\ast}$,
 $\Psi_{N^{\prime\ast}}$ and $A_\mu$ stand for the 
pion, nucleon, $\Delta$, $N^\ast$,
 $N^{\prime\ast}$ and photon fields, respectively 
; $N^\ast$ and $N^{\prime\ast}$ stand for the $N^\ast$(1440) and $N^\ast$(1520); 
resonances
$m$ and $\mu$ are the nucleon
and the pion masses; $\vec{\sigma}$ and $\vec{\tau}$ 
are the spin and isospin 1/2
operators; $\vec{S^\dagger}$ and $\vec{T}^\dagger$ 
are the transition spin and isospin
operators from 1/2 to 3/2 with the normalization
\begin{equation}
\langle\frac{3}{2},M|S_\nu^\dagger|\frac{1}{2},m\rangle = C (\frac{1}{2},1,
\frac{3}{2};m,\nu,M)
\end{equation}
with $\nu$ in spherical base, and the same for $T^\dagger$. The operators 
$\vec{S}_\Delta$ and $\vec{T}_\Delta$ are the ordinary spin and isospin matrices
for the a spin and isospin 3/2 object.
For the pion fields we used the Bjorken and Drell convention :
\begin{equation}
\phi_+ = \frac{1}{\sqrt{2}}(\phi_1 - i\phi_2)\hspace{0.5cm} 
destroys \hspace{0.4cm}\pi^+, creates \hspace{0.4cm}\pi^-
\end{equation}
\begin{equation}
\phi_- = \frac{1}{\sqrt{2}}(\phi_1 + i\phi_2)\hspace{0.5cm}
 destroys \hspace{0.4cm} \pi^-, creates \hspace{0.4cm} \pi^+
\end{equation}
\begin{equation}
\phi_0 =\phi_3\hspace{0.5cm} destroys \hspace{0.4cm} \pi^0, creates 
\hspace{0.4cm} \pi^0  
\end{equation}
Hence the $|\pi^+\rangle$ state corresponds to 
$- |1 1\rangle$ in isospin base.
\\
In all formulae we have assumed that $\sigma^i \equiv\sigma_i$, $S^i\equiv
S_i$, $T^i\equiv T_i$ are Euclidean vectors. However for $\partial_i$, $A_i$,
 $p_i$, etc, we have respected their covariant meaning.
\\

{\Large{\bf A2. Feynman Rules.}}
\\

Here we write the Feynman rules for the different vertices including
already the electromagnetic form factors. We assumed the photon
with momenta $q$ as an incoming particle while the pion with momentum  
$k$ is an outgoing particle in all vertices. The momentum $p$, $p^\prime$ are
those of the baryonic states just before and after the photon absorption vertex
(or pion production vertex in eq. (42)).

\begin{equation}
V^\mu_{\gamma N N} = - ie\left\{\begin{array}{c} 
F_1^N(q^2) \\ F_1^N(q^2)[\frac{\vec{p}+\vec{p}\, {^\prime}}{2m}] + i 
\frac{\vec{\sigma}\times\vec{q}}{2m}G_M^N(q^2) \end{array} \right\}
\end{equation}
\begin{equation}
V^\mu_{\gamma N \Delta} = \sqrt{\frac{2}{3}}\frac{f_\gamma(q^2)}{m_\pi}
\frac{\sqrt{s}}{m_\Delta}
\left\{\begin{array}{c} \frac{\vec{p}_\Delta}{\sqrt{s}}
(\vec{S^\dagger}\times\vec{q}) \\
\frac{p^0_\Delta}{\sqrt{s}} [\vec{S^\dagger}\times(\vec{q}-\frac{q^0}
{p^0_\Delta}\vec{p_\Delta})] \end{array} \right\}
\end{equation}
\begin{equation}
V_{\pi N \Delta} = -\frac{f^\ast}{\mu}\vec{S^\dagger}\cdot(\vec{k}-
\frac{k^0}{\sqrt{s}}\vec{p_\Delta})T^{\lambda\dagger}
\end{equation}
\begin{equation}
V_{\pi N N} = - \frac{f}{\mu}(\vec{\sigma}\vec{k}-k^0\frac{\vec{\sigma}
(\vec{p}+\vec{p}\, ^{\prime})}{2m})\tau^\lambda
\end{equation}
%\begin{equation}
%V_{N^\ast N \gamma} = \frac{\tilde{f}_\gamma^N}{\mu}
%\left\{\begin{array}{c} 0 \\
%(\vec{\sigma}\times
%\vec{q}) \end{array} \right\} 
%Scalar part:
\begin{equation}
V^0_{N^\ast N \gamma} = 
i\frac{\vec{q}\, ^2}{2m} F_2(q^2)
-i\vec{q}\, ^2 (1 + \frac{q^0}{2m}) F_1(q^2)
\end{equation}

\begin{eqnarray}
V^i_{N^\ast N \gamma}=
F_2(q^2)[i\vec{q}\,\frac{q^0}{2m} + (\vec{\sigma}\times\vec{q})
(1+\frac{q^0}{2m})]
\\ \nonumber & & 
\hspace{-6.5cm} 
-F_1(q^2)[i\vec{q}q^0(1+\frac{q^0}{2m})+q^2\frac{1}{2m}
(\vec{\sigma}\times\vec{q})]
\end{eqnarray}

\begin{equation}
V_{N^\ast \Delta \pi} = - \frac{g_{N^\ast \Delta \pi}}{\mu}\vec{S}^\dagger
\cdot\vec{k}T^{\lambda\dagger}
\end{equation}
\begin{equation}
V_{\Delta \Delta \pi} = - \frac{f_\Delta}{\mu}\vec{S_\Delta}\cdot\vec{k}\, 
 T_\Delta^\lambda
\end{equation}
\begin{equation}
V_{N^{\ast\prime} \Delta \pi} = -(\tilde{f}_{N^{\ast\prime} \Delta \pi}+
\frac{\tilde{g}_{N^{\ast\prime}
\Delta \pi}}{\mu^2}\vec{S^\dagger}\cdot\vec{k}\vec{S}\cdot\vec{k})
T^{\lambda\dagger}
\end{equation}
\begin{equation}
V^\mu_{\gamma \Delta \Delta} = - i\left\{\begin{array}{c}
e_\Delta F_1^\Delta(q^2) \\ e_\Delta 
F_1^\Delta(q^2)[\frac{\vec{p}+\vec{p^\prime}}
{2m_\Delta}] + i
\frac{\vec{S_\Delta}\times\vec{q}}{3m}eG_M^\Delta(q^2) \end{array} \right\}
\end{equation}
\begin{equation}
V^\mu_{\pi\pi\gamma} = -iq_\pi(k^\mu +k^{\prime\mu}) F_{\gamma\pi\pi}(q^2)
\end{equation}
\begin{equation}
V^\mu_{\Delta N \gamma \pi} = -q_\pi\frac{f^\ast}{m_\pi}T^{\lambda\dagger}
\left\{\begin{array}{c}
\, \vec{S}^\dagger\, \frac{\vec{p}_\Delta}{\sqrt{s}} \\ \vec{S}^\dagger 
\end{array} \right\} F_c(q^2)
\end{equation}

\begin{equation}
V^0_{\gamma N N^{\prime\ast}}=i(G_1(q^2) + G_2(q^2) p^{\prime 0} +
 G_3(q^2) q^0)\vec{S}^\dagger
\cdot\vec{q}
\end{equation}

\begin{eqnarray}
V^i_{\gamma N N^{\prime\ast}}=-i
[(\frac{G_1(q^2)}{2m} - G_3(q^2))(\vec{S}^\dagger\cdot\vec{q})\, \vec{q} - 
iG_1(q^2)\frac{\vec{S}^\dagger\cdot\vec{q}}{2m}(\vec{\sigma}\times\vec{q})
\\ 
\nonumber & & 
\hspace{-10cm} 
-\vec{S}^\dagger \{G_1(q^2)(q^0+\frac{\vec{q}\, ^2}{2m}) + 
G_2(q^2) p^{\prime 0} q^0 + G_3(q^2)
q^2\}]
\end{eqnarray}
\\

\newpage

{\Large{\bf A3. Coupling and form factors}}
\\

%\begin{tabular}{|r|l|c|c|c|c|c|c|c|} \hline
%Reaction & D1 & D2 & D3 & D4 & D5 & D6 & D7 & D8\\
%\hline\hline
%$\gamma_v p \rightarrow \pi^-\Delta^{++}$ & -i/3 & -i/3 & 0 & i/3 & 1 & 
%i/3 & i & -1 \\
%\hline
%$\gamma_v p \rightarrow \pi^+\Delta^0$ & i/9 & i/9 & -2i/9 & i/9 & -2/3 &
% 0 & i/3 & -1/3\\
%\hline
%\end{tabular}
%
%\vspace{0.3cm}
%
%Table A3: Coefficients of the amplitudes for the $\Delta^{++}$ and 
%$\Delta^0$ ($\Delta^0\rightarrow\pi^{-} p$) reactions, accounting
%for isospin and constant factors. 

%\vspace{0.3cm}
{\bf Coupling constants :} 
\\*[0.5cm]
$f = 1$\hspace{2cm} $f^\ast = 2.13$ 
\\*[0.5cm] 
$f_{\Delta N \gamma} = 0.122$\hspace{1cm} $f_\Delta = 0.802$
\\*[0.5cm]
$e = 0.3027$ 
\\*[0.5cm]
$\tilde{f}_{N^{\ast\prime}\Delta\pi} = -0.911$\hspace{2cm}  $\tilde{g}_{N^{\ast\prime}
\Delta \pi} = 0.552$
\\*[0.5cm]
$g_{N^\ast\Delta\pi} = 2.07$
\\*[0.5cm]
%$\tilde{f}_\gamma = \left\{\begin{array}{c}
%0.0173 \hspace{0.25cm}proton \\ -0.0112 \hspace{0.25cm}neutron 

%\end{array} \right\}$
%\\*[0.5cm]
%$\tilde{g}_\gamma = \left\{\begin{array}{c}
%0.108\hspace{0.25cm} proton \\ -0.129\hspace{0.25cm}neutron 
%\end{array} \right\}$
%\\*[0.5cm]
%$\tilde{g}_\sigma = \left\{\begin{array}{c}
%-0.049 \hspace{0.25cm}proton \\ -0.0073\hspace{0.25cm} neutron 
%\end{array} \right\}$
%\\*[0.5cm]
{\bf Form factors :}
\\*[0.5cm]

For the off-shell pions we use a form factor of the monopole type :
\begin{equation}
F_\pi(p^2) = \frac{\Lambda_\pi^2 - \mu^2}{\Lambda_\pi^2-p^2}\hspace{1.5cm}
;\, \Lambda_\pi\sim 1250\, MeV
\end{equation}

The Sachs's form factors are given by 
\begin{equation}
G_M^N(q^2) = \frac{\mu_N}{(1 - \frac{q^2}{\Lambda^2})^2}\hspace{1.5cm}
;\, G_E^N(q^2) = \frac{1}{(1 - \frac{q^2}{\Lambda^2})^2}
\end{equation}
with $\Lambda^2 = 0.71$ $GeV^2$; $\mu_p = 2.793$; $\mu_n = -1.913$. 
\\

The relation
between $F_1^p(q^2)$   (Dirac's form factor) and $G_E^p(q^2)$ is :

\begin{equation}
F_1^p(q^2) = G_E^p(q^2)\frac{(1 -\frac{q^2\mu_p}{4m_N^2})}{(1 -
\frac{q^2}{4m_N^2})}
\end{equation}
and $F_1^n = 0.$
\\

For the delta resonance we use
\begin{equation}
F_1^\Delta = F_1^p(q^2)
\end{equation}
\begin{equation}
G_M^\Delta(q^2) = \frac{\mu_\Delta}{(1 - \frac{q^2}{\Lambda^2})^2}
\end{equation}

In the case of the $\Delta^{++}$ we make use of the experimental value
$\mu_{\Delta^{++}} = 1.62$ $\mu_p$ and for the other charge states we make use 
of the ratio ($e > 0$):

\begin{equation}
\frac{\mu_\Delta}{\mu_p}=\frac{e_\Delta}{e}
\end{equation}

For the $\gamma \pi\pi$ vertex in diagram D2, we take the form factor:

\begin{equation}
F_{\gamma\pi\pi}(q^2) = \frac{1}{(1-\frac{q^2}{\lambda_\pi^2})}
\end{equation}
with $\lambda_\pi^2 = 0.5$ $GeV^2$.
\\

The axial nucleon form 
factor is given by:
\begin{equation}
F_A(q^2) = \frac{1}{(1-\frac{q^2}{M_A^2})^2}
\end{equation}
with $M_A$ = 1.16 GeV. This form factor is used for the contact form factor,
$F_c(q^2)$, in some of the results reported here. Some others results
use $F_c(q^2)$ = $F_{\gamma\pi\pi}(q^2)$.
\\

The form factor for the $\gamma \Delta N$ transition is taken as:
\begin{equation}
f_\gamma(q^2) = f_\gamma(0)\frac{(1 - \frac{q^2}{(m_\Delta + m)^2})}
{(1 -\frac{q^2\mu_p}{4m_N^2})}\, \frac{G_M^p(q^2)}{\mu_p}\, 
\frac{(m_\Delta + m)^2}
{(m_\Delta + m)^2 - q^2}
\end{equation}
where $f_\gamma(0)=0.122$ 
\\

{\Large{\bf A4. Amplitudes for the reaction}}
\\

In this appendix we write the explicit expressions for the amplitudes of the
Feynman diagrams used in the model. The isospin coefficients and some constant
factors are collected in the coefficients $C$ which are written in the table A4.
\\

\begin{tabular}{|r|l|c|c|c|c|c|c|c|} \hline
Reaction & D1 & D2 & D3 & D4 & D5 & D6 & D7 & D8\\
\hline\hline
$\gamma_v p \rightarrow \pi^-\Delta^{++}$ & -i/3 & -i/3 & 0 & i/3 & 1 & 
i/3 & i & -1 \\
\hline
$\gamma_v p \rightarrow \pi^+\Delta^0$ & i/9 & i/9 & -2i/9 & i/9 & -2/3 &
 0 & i/3 & -1/3\\
\hline
\end{tabular}

\vspace{0.3cm}

Table A4: Coefficients of the amplitudes for the $\Delta^{++}$ and 
$\Delta^0$ ($\Delta^0\rightarrow\pi^{-} p$) reactions, accounting
for isospin and constant factors.
\vspace{0.3cm}

In the following expressions $q$, $p_1$, $p_2$, $p_4$, and $p_5$ are the momentum of the 
photon,the incoming nucleon, the outgoing nucleon and the two pions :
\\*[0.5cm]
\begin{tabular}{|r|r|c|c|c|}\hline  
$\gamma$ & p  & $\pi^+$ & $\pi^-$ & p\\
\hline\hline
q & $p_1$ & $p_5$ & $p_4$ & $p_2$ \\
\hline
\end{tabular}
\\*[0.5cm]
\\

We write only the amplitude when the pion labelled $p_5$ is
emitted before the pion labelled $p_4$($\Delta^0$ case), except
in the case of the D6 where the only possibility is the $\Delta^{++}$
and the explicit amplitude for this case is written. We have also
evaluated the crossed diagrams when the pion labelled $p_5$ is emitted 
after the pion called $p_4$($\Delta^{++}$ case).
Such amplitudes are exactly the same than the others written before, but
exchanging the momenta $p_4$ and $p_5$ and changing some isospin coefficient.
 This latter change is taken into account by the factor $C$ written in table
 A4.

We should note that in the vertex  $\Delta N \pi$, when $\vec{p}_\Delta$ is
not zero, we must change $\vec{p}_\pi$ by $\vec{p}_\pi - \frac{p^0}
{\sqrt{s}}\vec{p}_\Delta$ for the final pion.

In the formulae, $D_\pi$, $G_\Delta$, $G_N$, $G_{N^\ast}$, $G_{N^{\prime\ast}}$ 
are the propagator of the pion, delta, nucleon, $N^\ast(1440)$,  
$N^\ast(1520)$ respectively. Expressions for them and for the width of the 
resonances can be found in [1, 2, 14].

\begin{eqnarray}
-i T_1^\mu & = & Ce(\frac{f^\ast}{\mu})^2
G_\Delta(p_2 + p_4) F_\pi((p_5-q)^2)
F_c(q^2)  \\ 
& & \times\left\{\begin{array}{c}
[2\vec{p}_4\cdot\vec{p}_5 - i(\vec{p}_4\times\vec{p}_5)\cdot\vec{\sigma}]
\frac{-1}{\sqrt{s_\Delta}}  \\ \nonumber
2\vec{p}_4 - i (\vec{\sigma}\times\vec{p}_4)
\end{array} \right\}
\end{eqnarray}
\begin{eqnarray}
-i T_2^\mu & = & Ce(\frac{f^\ast}{\mu})^2
G_\Delta(p_2 + p_4)D_\pi(p_5 - q) F_\pi((p_5-q)^2)
F_{\gamma\pi\pi}(q^2) \\ \nonumber
& & \times 
[2\vec{p}_4\cdot(\vec{p}_5-\vec{q}) - i(\vec{p}_4\times(\vec{p}_5-\vec{q}))
\cdot\vec{\sigma}] \\ \nonumber
& & \times\left\{\begin{array}{c}  
2p_5 - q \nonumber
\end{array} \right\}^\mu 
\end{eqnarray}
\begin{eqnarray}
-i T_3^\mu & = & C\frac{f}{\mu}\frac{f^\ast}{\mu}
\frac{f_\gamma(q^2)}{\mu}
G_N(p_2 + p_4)G_\Delta(p_1 + q) \\ 
& & \times\left\{\begin{array}{c}
[-2i(\vec{p}_4\times\vec{q})
-(\vec{\sigma}\cdot\vec{q})\vec{p}_4
+(\vec{p}_4\cdot
\vec{q})\vec{\sigma}]\frac{\vec{p}_\Delta}{m_\Delta}
 \\ \nonumber
[-2i(\vec{p}_4\times\vec{q}^{\, \prime} )
-(\vec{\sigma}\cdot\vec{q}^{\, \prime} )\vec{p}_4
+(\vec{p}_4\cdot
\vec{q}^{\, \prime} )\vec{\sigma}]\frac{p_\Delta^0}{m_\Delta}
\end{array} \right\} \\ \nonumber
& & \times [-p_5^0\frac{\vec{\sigma}(2\vec{p}_1-\vec{p}_5)}{2m} + 
\vec{\sigma}\vec{p_5}]
\end{eqnarray}
with $\vec{q}^{\, \prime} =
( \vec{q} - \frac{q^0}{p_\Delta^0}\vec{p_\Delta} )$
\begin{eqnarray}
-i T_4^\mu & = & C e(\frac{f^\ast}{\mu})^2 G_N(p_1 + k)G_\Delta(p_2 +
p_4) F_\pi((p_5-q)^2) \\ \nonumber
& & \times[2\vec{p}_4\cdot\vec{p}_5 - 
i(\vec{p}_4\times\vec{p}_5)\cdot\vec{\sigma}] \\ \nonumber
& & \times\left\{\begin{array}{c}
F_1^p(q^2) \\ \nonumber
F_1^p(q^2)[\frac{\vec{p}+\vec{p} ^{\, \prime}}{2m}] + iG_M^p(q^2)\frac{\vec{\sigma}
\times\vec{q}}{2m} \nonumber
\end{array} \right\} \\ \nonumber
\end{eqnarray}
\begin{equation}
-iT_5^0 = 0 \hspace{0.4cm} in \hspace{0.3cm}\gamma-p\hspace{0.3cm}CM
\hspace{0.3cm}frame
\end{equation}
\begin{eqnarray}
-i T_5^i & = & C \frac{f^\ast}{\mu}\frac{f_\Delta}{\mu}\frac{f_\gamma(q^2)}
{\mu}G_\Delta(p_2 + p_4)G_\Delta(p_1 + q) \\ \nonumber
& & \times [i\frac{5}{6}(\vec{p}_4\cdot\vec{q})\vec{p}_5 - 
i\frac{5}{6}(\vec{p}_5\cdot\vec{q})\vec{p}_4 -
\frac{1}{6}(\vec{p}_4\cdot\vec{p}_5)(\vec{\sigma}\times\vec{q})-\\ \nonumber
& & \frac{1}{6}(\vec{p}_4\cdot\vec{\sigma})(\vec{p}_5\times\vec{q}) +
\frac{2}{3}(\vec{p}_5\cdot\vec{\sigma})(\vec{p}_4\times\vec{q})] \nonumber
\end{eqnarray}
\begin{eqnarray}
-i T_6^0 & = & C(\frac{f^\ast}{\mu})^2G_\Delta(p_2 + p_5)G_\Delta(p_1 - p_4)
F_\pi((p_5-q)^2)
\{e_\Delta F_1^\Delta(q^2) \\ \nonumber
& & 
\times[2\vec{p}_5\cdot\vec{p}_4 
- i(\vec{p}_5\times\vec{p}_4)\cdot\vec{\sigma}]\} \\ \nonumber
\end{eqnarray}

\begin{eqnarray}
-i T_6^i & = & C(\frac{f^\ast}{\mu})^2G_\Delta(p_2 + p_5)G_\Delta(p_1 - p_4)
F_\pi((p_5-q)^2) \\ 
\nonumber
& & 
\times\{\frac{e_\Delta F_1^\Delta(q^2)}{2}
\frac{(\vec{p}_1-2\vec{p}_4)}{m_\Delta}
[2\vec{p}_5\cdot\vec{p}_4 
- i(\vec{p}_5\times\vec{p}_4)\cdot\vec{\sigma}]+ \\ \nonumber
& & i\frac{eG_M^\Delta(q^2)}{m}[i\frac{5}{6}(\vec{p}_4\cdot\vec{q})\vec{p}_5 -
i\frac{5}{6}(\vec{p}_5\cdot\vec{q})\vec{p}_4 -
\frac{1}{6}(\vec{p}_5\times\vec{q})(\vec{p}_4\cdot\vec{\sigma})-\\ \nonumber
& & \frac{1}{6}(\vec{p}_5\cdot\vec{\sigma})(\vec{p}_4\times\vec{q}) +
\frac{2}{3}(\vec{p}_5\cdot\vec{p}_4)(\vec{\sigma}\times\vec{q})]\} \nonumber
\end{eqnarray}
{\bf Amplitude of $N^{\prime\ast}$(1520):}
\\
Vector part :
\\
\begin{eqnarray}
-iT_7^i & = &
C\frac{f^\ast}{\mu}G_\Delta(p_2 + p_4)G_{N^{\prime\ast}}(p_1 + q) \\
\nonumber
& &
\times\vec{S}\cdot\vec{p}_4 [\tilde{f}_{N^{\prime\ast}\Delta\pi}
+ \frac{\tilde{g}_{N^{\prime\ast}\Delta\pi}}{\mu^2}
\vec{S}^\dagger\cdot\vec{p}_5\vec{S}\cdot\vec{p}_5] \\ 
\nonumber
 & & \times\{(\frac{G_1(q^2)}{2m} - G_3(q^2))(\vec{S}^\dagger\cdot\vec{q})\, 
 \vec{q} - 
i G_1(q^2)\frac{\vec{S}^\dagger\cdot\vec{q}}{2m}(\vec{\sigma}\times\vec{q}) \\
\nonumber
& & 
-\vec{S}^\dagger [G_1(q^2)(q^0+\frac{\vec{q}\, ^2}{2m}) + 
G_2(q^2) p^{\prime 0} q^0 + G_3(q^2)
q^2]\}
\end{eqnarray}
\\
scalar part:

\begin{eqnarray}
-iT_7^0 & = &
-C\frac{f^\ast}{\mu}G_\Delta(p_2 + p_4)G_{N^{\prime\ast}}(p_1 + q) \\
\nonumber
& &
\times\vec{S}\cdot\vec{p}_4 [\tilde{f}_{N^{\prime\ast}\Delta\pi}
+ \frac{\tilde{g}_{N^{\prime\ast}\Delta\pi}}{\mu^2}
\vec{S}^\dagger\cdot\vec{p}_5\vec{S}\cdot\vec{p}_5] \\ 
\nonumber
& &
\times [G_1(q^2) + G_2(q^2) p^{\prime 0} 
+ G_3(q^2) q^0] \vec{S}^\dagger\cdot\vec{q}
\nonumber
\end{eqnarray}  
{\bf Amplitude of $N^{\ast}$(1440):}
\\
Vector part :
\\
\begin{eqnarray}
-iT_8^i & = & C 
\frac{f^\ast}{\mu}\frac{g_{N^{\ast}\Delta\pi}}{\mu}
G_\Delta(p_2 + p_4) G_{N^\ast}(p_1 + q)\vec{S}\cdot
\vec{p}_4\vec{S}^\dagger\cdot\vec{p}_5 \\ 
& & \nonumber\times\
\{F_2(q^2)[i\vec{q}\,\frac{q^0}{2m}+(\vec{\sigma}\times\vec{q})
(1+\frac{q^0}{2m})]
\\ \nonumber & & 
-F_1(q^2)[i\vec{q}q^0(1+\frac{q^0}{2m})+q^2\frac{1}{2m}
(\vec{\sigma}\times\vec{q})]\}
\end{eqnarray}
\\
scalar part:

\begin{eqnarray}
-iT_8^0 & = & C 
\frac{f^\ast}{\mu}\frac{g_{N^{\ast}\Delta\pi}}{\mu}\
G_\Delta(p_2 + p_4) G_{N^\ast}(p_1 + q) \vec{S}\cdot
\vec{p}_4\vec{S}^\dagger\cdot\vec{p}_5\\ 
& & \nonumber \times\
\{i\frac{\vec{q}\, ^2}{2m} F_2(q^2)
-i\vec{q}\, ^2 (1 + \frac{q^0}{2m}) F_1(q^2)\}
\end{eqnarray}

{\Large{\bf A5. Amplitudes for the gauge invariant set within Berends formalism. }}
\\

The explicit amplitudes for the gauge invariant set of the 4 diagrams 
D1, D2, D4 and D6, in the
presence of different form factors are 
made according to eq. (17) of \cite{berends}. We implement them in our
formalism by making the following
substitutions in the amplitudes shown in the appendix A4.
The form factor in the zeroth component of each amplitude is changed as:
\\

 $T_1^\mu$:
 
\begin{equation}
\frac{\vec{S}^\dagger\cdot\vec{p}_\Delta}{\sqrt{s_\Delta}}F_c\rightarrow
\frac{\vec{S}^\dagger\cdot\vec{p}_\Delta}{\sqrt{s_\Delta}}F_c
-(F_c - 1) (\frac{\vec{S}^\dagger\cdot\vec{p}_\Delta}{\sqrt{s_\Delta}}q^0
-\vec{S}^\dagger\cdot\vec{q})\frac{q^0}{q^2}
\end{equation}

 $T_2^\mu$ :

\begin{equation}
F_{\gamma\pi\pi}\{2p_5 - q\}^0\rightarrow
F_{\gamma\pi\pi}\{2p_5 - q\}^0 + (F_{\gamma\pi\pi} - 1) D^{-1}_\pi\, 
\frac{q^0}{q^2}
\end{equation}

 $T_4^\mu$ :

\begin{equation}
F_1^p\rightarrow F_1^p +
(F_1^p - 1) G_N^{-1}\, \frac{q^0}{q^2}
\end{equation} 

 $T_6^\mu$ :
\begin{equation}
F_1^{\Delta^{++}}\rightarrow
F_1^{\Delta^{++}} + (F_1^{\Delta^{++}} - 1) G_\Delta^{-1}\, \frac{q^0}{q^2}
\end{equation}
where $G_N$, $G_\Delta$ are the non relativistic propagator of the nucleon
and the $\Delta$ resonance and $D_\pi$ the ordinary relativistic pion propagator.
\\

{\Large{\bf A6. Miscellaneous Formulae. }}
\\

In order to obtain our amplitudes we have employed some useful relations:
\begin{equation}
\sum_M S_i|M\rangle\langle M|S_j^\dagger = \frac{2}{3}\delta_{ij} -
\frac{i}{3}\epsilon_{ijk}\sigma_k = \delta_{ij} - \frac{1}{3}\sigma_i\sigma_j
\end{equation}
\begin{eqnarray}
\sum_{MM^\prime} S_i|M\rangle\langle M|S_{\Delta j}|M^{\prime}\rangle\langle|
M^{\prime}|S_k^\dagger & = & \frac{5}{6}i\epsilon_{ijk} -\frac{1}{6}
\delta_{ij}
\sigma_k + \\ \nonumber &  &\frac{2}{3}\delta_{ik}\sigma_j - \frac{1}{6}
\delta_{jk}\sigma_i \nonumber
\end{eqnarray}

From eqs. (78) and (79). we can prove the following relations,which are used in
the calculation of the amplitudes:
\begin{equation}
\vec{S}\cdot\vec{p}\vec{S^\dagger}\cdot\vec{q} = \frac{1}{3}[2\vec{p}
\cdot\vec{q} - i(\vec{p}\times\vec{q})\cdot\vec{\sigma}]
\end{equation}
\begin{eqnarray}
\vec{S}\cdot\vec{p}(\vec{S_\Delta}\times\vec{k})\cdot\vec{\epsilon}
\vec{S^\dagger}\cdot
\vec{q} &  =  & [\frac{5}{6}i\vec{p}\cdot\vec{\epsilon}\vec{q}\cdot\vec{k} - 
\frac{5}{6}i\vec{p}\cdot\vec{k}\vec{q}\cdot\vec{\epsilon}- 
\\ \nonumber &  & \frac{1}{6}
(\vec{p}\times\vec{k})\cdot\vec{\epsilon}\vec{\sigma}\cdot\vec{q} -
\frac{1}{6}\vec{\sigma}\cdot\vec{p}(\vec{q}\times\vec{k})\cdot
\vec{\epsilon} + \\ \nonumber &  &
\frac{2}{3}\vec{p}\cdot\vec{q}(\vec{\sigma}\times\vec{k})
\cdot\vec{\epsilon}] \nonumber
\end{eqnarray}
where we have omitted the sum over intermediate states.
\\

\end{document}